\documentclass[preprint, sort&compress]{elsarticle}
\usepackage{graphicx}
\usepackage{multirow}
\usepackage[usenames]{color}
\graphicspath{{ps}}
\usepackage{pstricks}

\newcommand{\1}{\overline{B}{}_s^0 \to \Lambda_c^+ \overline{\Lambda} \pi^-}
\newcommand{\Br}[1]{{\cal B}_{#1}}
\newcommand{\Bs}[1]{B_s^{#1}}
\newcommand{\Bsbar}[1]{\overline{B}{}_s^{#1}}
\newcommand{\DE}{\Delta E}
\newcommand{\gev}{{\rm~GeV}}
\newcommand{\K}{{\it K}}
\newcommand{\Mbc}{M_{\rm bc}}
\newcommand{\mev}{{\rm~MeV}}
\newcommand{\pion}{$\pi$}
\newcommand{\p}{{\it p}}
\newcommand{\thrust}{{\rm \left| \cos \Theta_{thrust} \right|}}

\newlength{\first}
\newlength{\second}

\begin{document}

\settowidth{\first}{BELLE Preprint 2013-4}

\setlength{\second}{\textwidth}
\addtolength{\second}{-\first}

\begin{tabular*}{\textwidth}{p{\second} p{\first}}
    & BELLE Preprint 2013-4 \\
    & KEK Preprint 2012-48 \\
\end{tabular*}

\begin{frontmatter}

   \title{Evidence for $\1$}

   \author[ITEP]{Belle Collaboration \\ \quad \\
                    E.~Solovieva} 
      \author[ITEP]{R.~Chistov} 
      \author[KEK]{I.~Adachi} 
      \author[PNNL]{D.~M.~Asner} 
      \author[ITEP]{T.~Aushev} 
      \author[Sydney]{A.~M.~Bakich} 
      \author[Panjab]{A.~Bala} 
      \author[Nara]{V.~Bhardwaj} 
      \author[IITG]{B.~Bhuyan} 
      \author[Nara]{M.~Bischofberger} 
% ...

% ...
\address[KEK]{High Energy Accelerator Research Organization (KEK), Tsukuba 305-0801, Japan}
% ...
\address[IITG]{Indian Institute of Technology Guwahati, Assam 781039, India}
% ...
\address[ITEP]{Institute for Theoretical and Experimental Physics, Moscow 117218, Russian Federation}
% ...
\address[Nara]{Nara Women's University, Nara 630-8506, Japan}
% ...
\address[PNNL]{Pacific Northwest National Laboratory, Richland, WA 99352, USA}
   \address[Panjab]{Panjab University, Chandigarh 160014, India}
% ...
\address[Sydney]{School of Physics, University of Sydney, NSW 2006, Australia}
% ...

   \begin{abstract}

Using 121.4 fb$^{-1}$ of data collected with the Belle detector at the $\Upsilon (5S)$ resonance at the KEKB asymmetric-energy $e^+e^-$ collider, we report evidence for the $\1$ decay mode with a measured  branching fraction $\left(3.6 \pm 1.1[{\rm stat.}] \right.$ $\left.{+0.3 \atop -0.5}[{\rm syst.}] \pm 0.9[\Lambda_c^+] \pm 0.7[N_{\Bsbar{0}}] \right) \times 10^{-4}$ and a significance of 4.4 standard deviations. This is the first evidence for a baryonic $\Bs{0}$ decay.

   \end{abstract}

\end{frontmatter}

\section{Introduction}

Results on baryonic $B$-meson decays obtained by Belle~\cite{Abe2, Abe234, Abe3, Gabyshev, Chen, Abe4} and BaBar~\cite{Aubert2, Aubert23, Aubert3, Aubert34, Aubert345, Aubert4} have increased experimental and theoretical interest in such processes~\cite{Suzuki}. $B$-meson decay modes with two-~\cite{Abe2, Abe234, Aubert2, Aubert23}, three-~\cite{Abe234, Abe3, Gabyshev, Chen, Aubert23, Aubert3, Aubert34, Aubert345} and even four-~\cite{Abe234, Abe4, Aubert34, Aubert345, Aubert4} and five-body~\cite{Aubert345} final states have been observed. The measured branching fractions clearly follow a hierarchy that depends on the final-state multiplicity: two-body channels have smaller branching fractions compared to multi-body ones. In addition, most three-body baryonic $B$-meson decays have a near-threshold peak in the invariant baryon-antibaryon mass spectrum. This effect was investigated in Ref.~\cite{Hou}. In this Letter, we report the first evidence for the $\1$ decay and compare the measured branching fraction with that for a similar channel, $B^- \to \Lambda_c^+ \overline{p} \pi^-$~\cite{Gabyshev}, where the $s$-quark of the decay under study here is replaced by a $u$-quark~\cite{CC}.

\section{Data Sample and the Belle Detector}

The data for this analysis were taken with the Belle detector at the $e^+e^-$ asymmetric-energy collider KEKB~\cite{KEKB} at the $\Upsilon (5S)$ resonance. The integrated luminosity of the sample is 121.4 fb$^{-1}$ and corresponds to $\left(7.1 \pm 1.3 \right) \times 10^6$ $\Bs{0} \Bsbar{0}$ meson pairs~\cite{Esen} produced in three $\Upsilon (5S)$ decay channels: $\Bs{*0} \Bsbar{*0}$, $\Bs{*0} \Bsbar{0}$, and $\Bs{0} \Bsbar{0}$.

The Belle detector is a large-solid-angle magnetic spectrometer consisting of a silicon vertex detector (SVD), a 50-layer central drift chamber (CDC), an array of aerogel threshold Cherenkov counters (ACC), a barrel-like arrangement of time-of-flight scintillation counters (TOF), and an electromagnetic calorimeter (ECL) comprised of CsI(Tl) crystals located inside a superconducting solenoid coil providing a 1.5-T magnetic field. An iron flux return located outside the coil is instrumented to detect $K_L^0$ mesons and identify muons (KLM). The detector is described in detail elsewhere~\cite{Abashian}.

\section{Selection Criteria}

We use selection requirements previously used for baryonic $B$-meson decay analyses~\cite{Chen}. Charged tracks, except those from $K_S^0$ and $\Lambda$, are required to originate within 0.25 cm in the radial direction and within 1 cm along the beam direction from the interaction point (IP). We distinguish a charged particle of type $A$ from one of type $B$ ($A$ and $B$ being \pion, \K~or \p) based on likelihood values ${\cal L}(A)$ and ${\cal L}(B)$ derived from the TOF, ACC, and {\it dE/dx} measurements in the CDC.

$K_S^0$ mesons ($\Lambda$ hyperons) are reconstructed in the $K_S^0 \to \pi^+ \pi^-$ ($\Lambda \to p \pi^-$) decay mode by fitting the pion (\p{} and \pion{}) tracks to a common vertex, demanding an invariant mass in an interval of $\pm 10 \mev/c^2~[\approx 3 \sigma]~(\pm 4 \mev/c^2~[\approx 3 \sigma])$ around the nominal $\K_S^0$ ($\Lambda$) mass value~\cite{PDG} and applying the following requirements:

\begin{itemize}
   \item the distance of closest approach between daughter particles at the decay vertex should be less than 3 cm;
   \item the distance between the vertex position and IP in the plane transverse to the beam direction should be greater than 0.01 cm;
   \item the angle $\alpha$ between the $\K_S^0$ ($\Lambda$) momentum vector and the vector pointing from the IP to the $\K_S^0$ ($\Lambda$) decay vertex, measured in the plane transverse to the beam direction, should satisfy $\cos \alpha > 0.99$;
   \item the vertex fit should have $\chi^2 < 100~(10)$ for $K_S^0$ ($\Lambda$).
\end{itemize}

A sample of $\Lambda_c^+$ hyperons is reconstructed in the $\Lambda_c^+ \to p K^- \pi^+$, $\Lambda_c^+ \to p K_S^0$, and $\Lambda_c^+ \to \Lambda \pi^+$ decay modes. We apply a mass requirement on the reconstructed $\Lambda_c^+$ candidates, demanding the invariant mass be within the $10 \mev/c^2$ ($\approx 3 \sigma$) interval around the nominal mass value~\cite{PDG}.

\section{$\Bs{0}$-meson Reconstruction}

We fit the $\Lambda_c^+$ and $\overline{\Lambda}$ momentum vectors and the \pion{} track to a common $\Bs{0}$ vertex. To reject backgrounds including displaced tracks (e.g., unreconstructed $K_S^0$ and $\Lambda$ decays), we impose a loose requirement on the vertex-fit quality for the $\Bs{0}$. Signal candidates are identified by two kinematic variables computed in the $\Upsilon(5S)$ rest frame: the beam-energy-constrained mass $\Mbc = \sqrt{E_{\rm beam}^2 - p_{\Bs{0}}^2}$ and the energy difference $\DE = E_{\Bs{0}} - E_{\rm beam}$, where $E_{\rm beam}$ is the beam energy, and $E_{\Bs{0}}$ and $p_{\Bs{0}}$ are the energy and momentum, respectively, of the reconstructed $\Bs{0}$ candidate. For the $\Upsilon(5S) \to \Bs{0} \Bsbar{0}$ production channel, signal events correspond to a cluster at $(m_{\Bs{0}}$, 0) in the $\Mbc$ vs. $\DE$ plane. For the $\Upsilon(5S) \to \Bs{*0} \Bsbar{0}$ [$\Bs{*0} \Bsbar{*0}$] channel, the photon from the $\Bs{*0} \to \Bs{0} \gamma$ is not reconstructed and so signal events cluster at $( (m_{\Bs{0}} + m_{\Bs{*0}})/2$, $(m_{\Bs{0}} - m_{\Bs{*0}})/2 )$ $\left[ (m_{\Bs{*0}},~ m_{\Bs{0}} - m_{\Bs{*0}}) \right]$ in the $\Mbc$ vs. $\DE$ plane. We retain $\Bs{0}$ meson candidates with $\Mbc > 5.3 \gev/c^2$ and $\left| \DE\right| < 0.3 \gev$ for further analysis.

To suppress $e^+e^- \to c \overline{c}$ background, we require that the ratio $R_2$ of the second and zeroth Fox-Wolfram moments~\cite{Fox} be less than 0.5. We also specify that the angle $\rm \Theta_{thrust}$ between the thrust axis of the $\Bs{0}$ candidate in the $\Upsilon(5S)$ frame and the thrust axis of the rest of the event satisfies $\thrust{} < 0.85$. The mass window of the $\Lambda_c^+$ candidate, $R_2$ and $\rm \Theta_{thrust}$ requirements are optimized by maximizing a figure of merit (FOM) $N_{\rm sig}/\sqrt{N_{\rm bkgd}}$, where $N_{\rm sig}$ is the expected number of signal events from Monte Carlo simulation and $N_{\rm bkgd}$ is the expected number of background events estimated from the $\DE$ sidebands in the data.

Signal Monte Carlo samples of 120000 events each for different $\Bs{0}$-meson production modes and $\Lambda_c^+$ decay channels are used to evaluate the response of the detector and determine its efficiency. Events are generated using the EvtGen program; the detector response is simulated with GEANT~\cite{Lange}. We model the $\1$ decay according to phase-space hypothesis.

\section{Fit Procedure and Results}

We apply an unbinned extended maximum likelihood fit simultaneously to the three two-dimensional $\Mbc$ vs. $\DE$ spectra, corresponding to different $\Lambda_c^+$ subchannels. Signal and background distributions are parameterized separately for all subchannels, taking each function to be the product of shapes in $\Mbc$ and $\DE$. For the signal the linear correlation between $\Mbc$ and $\DE$ is less than 0.002, while the background correlation does not exceed 0.005.

The contribution of the $\Bs{0}$ production channel $C$ ($C$ being $\Bs{*0} \Bsbar{*0}$, $\Bs{*0} \Bsbar{0}$ or $\Bs{0} \Bsbar{0}$) is parameterized by a two-dimensional Gaussian with parameters determined from the Monte Carlo simulation. The typical resolution in $\Mbc$ is $3.6 \mev/c^2$ and in $\DE$ is between $8.8 \mev$ and $9.9 \mev$. The number of signal events for the channel $C$ is written as:
\begin{eqnarray}
   \label{normalization}
   N_C^{pK\pi} &=& N_{\Bsbar{0}} \, f_C \, \Br{\1} \, \Br{\Lambda_c^+ \to pK^- \pi^+} \, \Br{\Lambda \to p \pi^-} \, \epsilon_C^{pK\pi} \nonumber \\
   N_C^{pK_S^0} &=& N_{\Bsbar{0}} \, f_C \, \Br{\1} \, \Br{\Lambda_c^+ \to p K_S^0} \, \Br{K_S^0 \to \pi^+ \pi^-} \, \Br{\Lambda \to p \pi^-} \, \epsilon_C^{pK_S^0} \\
   N_C^{\Lambda\pi} &=& N_{\Bsbar{0}} \, f_C \, \Br{\1} \, \Br{\Lambda_c^+ \to \Lambda \pi^+} \, {\cal B}^2_{\Lambda \to p \pi^-} \, \epsilon_C^{\Lambda\pi}, \nonumber
\end{eqnarray}
where $f_C$ is the probability for the $\Bs{0}$-meson to be produced through the channel $C$ and $\epsilon$ is the reconstruction efficiency that is determined from the Monte Carlo simulation. For the fractions $f_C$, we use the following values~\cite{Esen}: $f_{\Bs{*0} \Bsbar{*0}} = (87.0 \pm 1.7)\%$, $f_{\Bs{*0} \Bsbar{0}} = (7.3 \pm 1.4)\%$, and $f_{\Bs{0} \Bsbar{0}} = 1 - f_{\Bs{*0} \Bsbar{*0}} - f_{\Bs{*0} \Bsbar{0}}$. The $\1$ branching fraction is a common parameter shared among subchannels, while the world average values~\cite{PDG} are used for the intermediate branching fractions. The average reconstruction efficiency is found to be 12.5\% for the $pK\pi$, 5.9\% for the $pK_S^0$, and 8.7\% for the $\Lambda\pi$ subchannel. 

Background shapes are described with an ARGUS threshold function~\cite{ARGUS} in $\Mbc$ and a linear function in $\DE$. We exclude the $\DE < -150 \mev$ region from the fit to avoid contributions from possible $\Bsbar{0} \to \Lambda_c^+ \overline{\Lambda} \pi^- \pi^0$ decays, where the $\pi^0$ is not reconstructed. This cutoff is verified by Monte Carlo simulation of $\Bsbar{0} \to \Lambda_c^+ \overline{\Lambda} \pi^- \pi^0$ and $\Bsbar{0} \to \Lambda_c^+ \overline{\Lambda} \rho^-$ decays.

\begin{figure}[t]
   \centering
   \includegraphics[width = 0.33\textwidth]{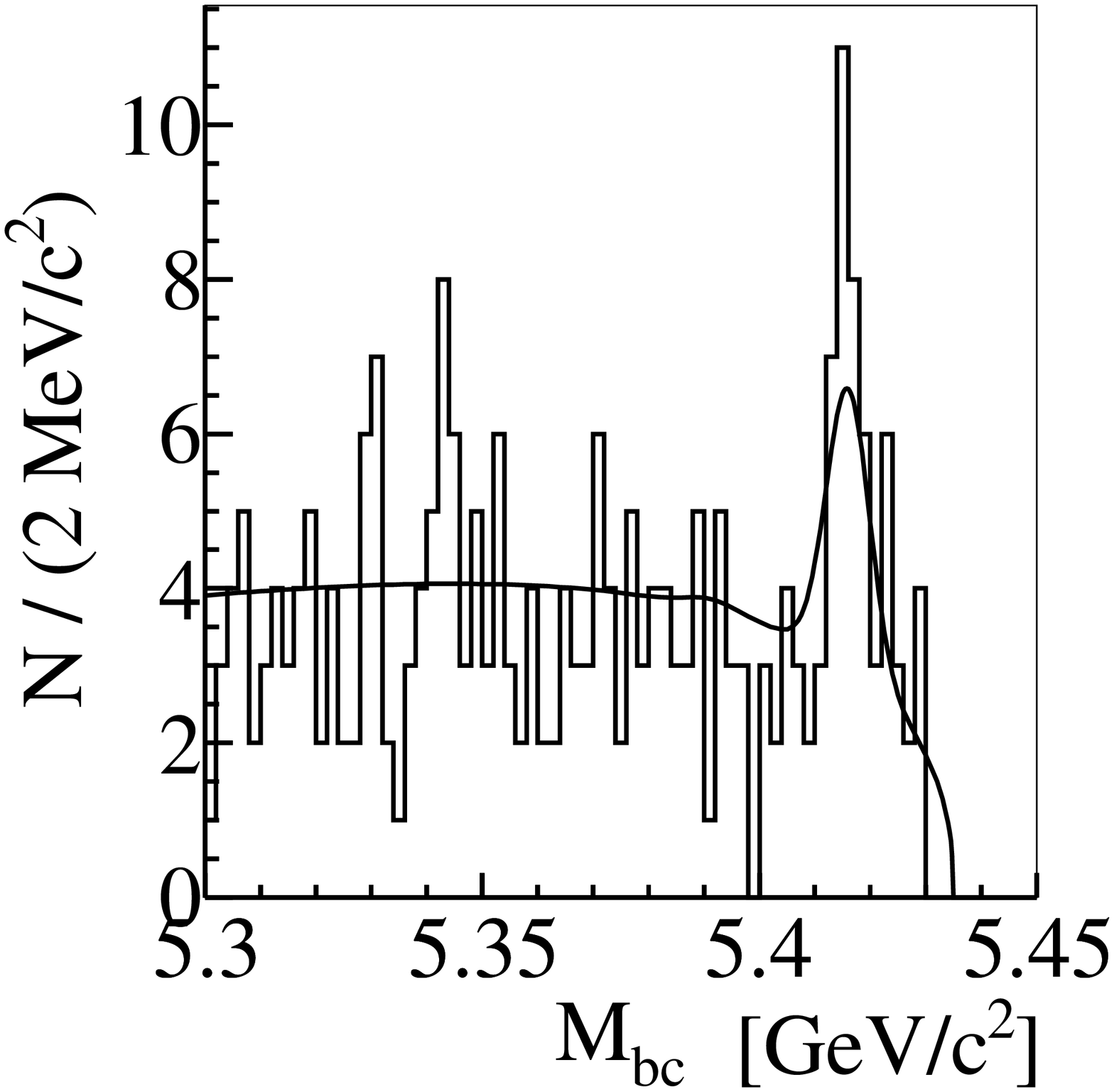}
   \includegraphics[width = 0.33\textwidth]{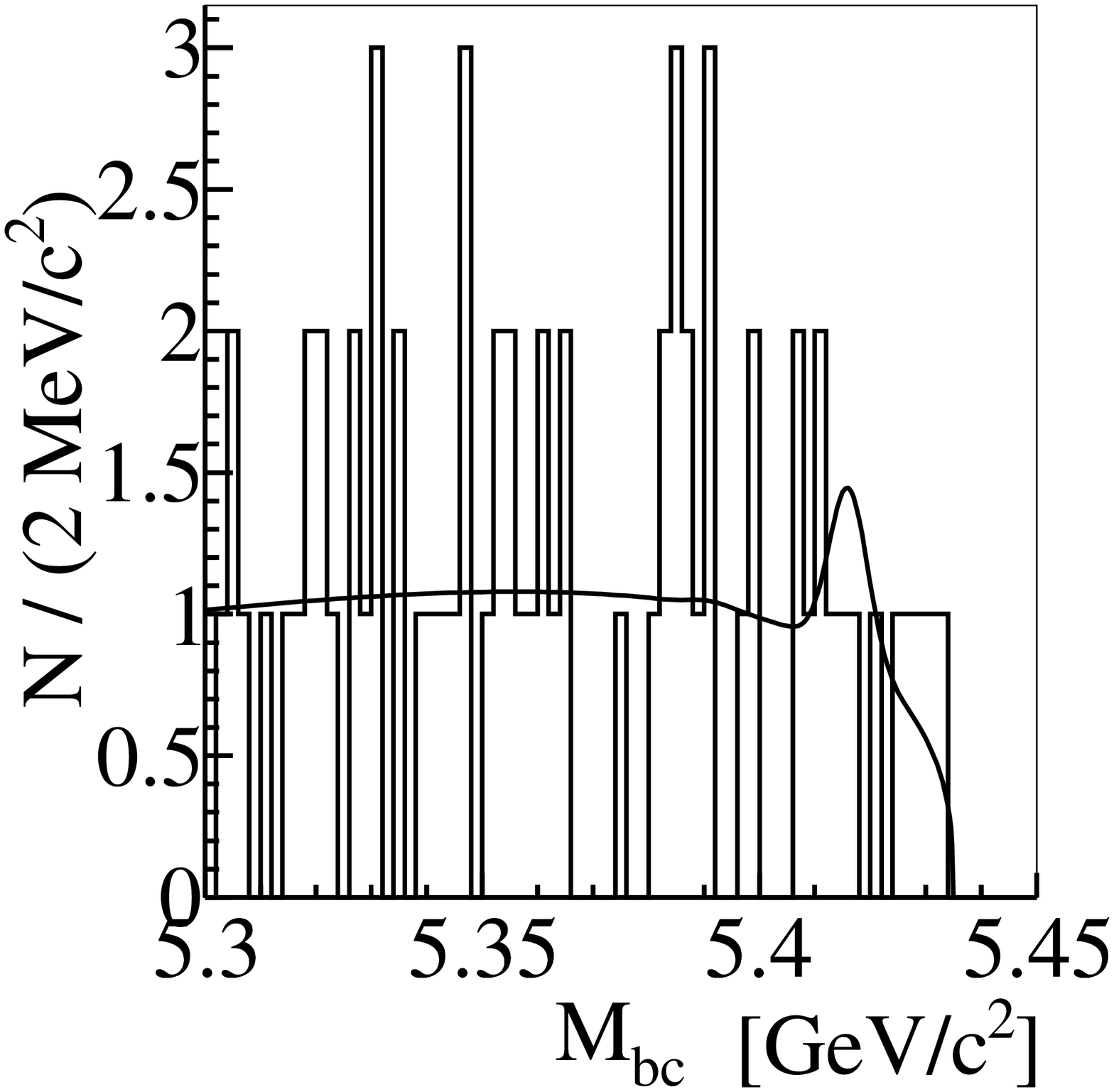}
   \put(-27,95){}
   \includegraphics[width = 0.33\textwidth]{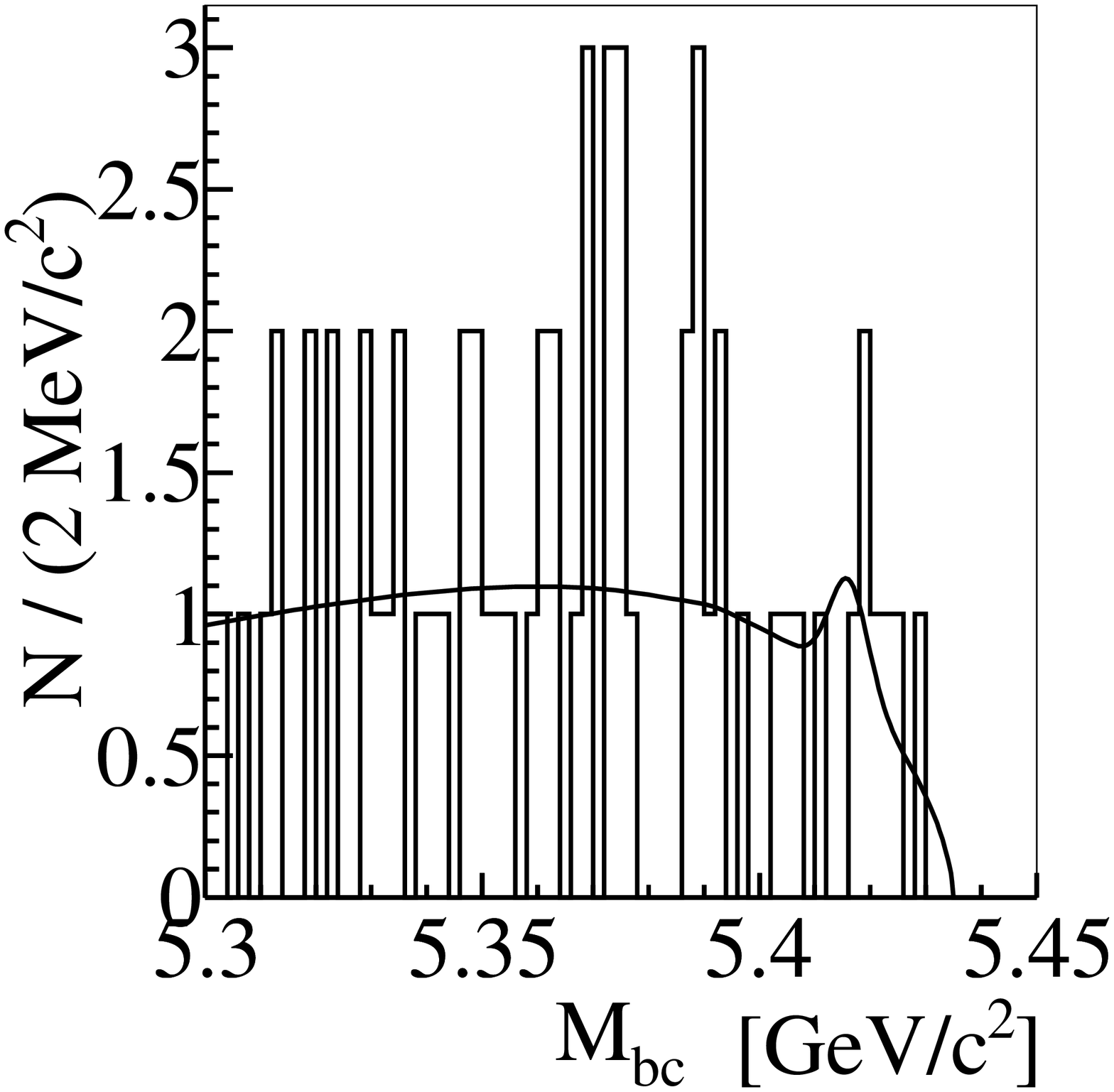} \\
   \includegraphics[width = 0.33\textwidth]{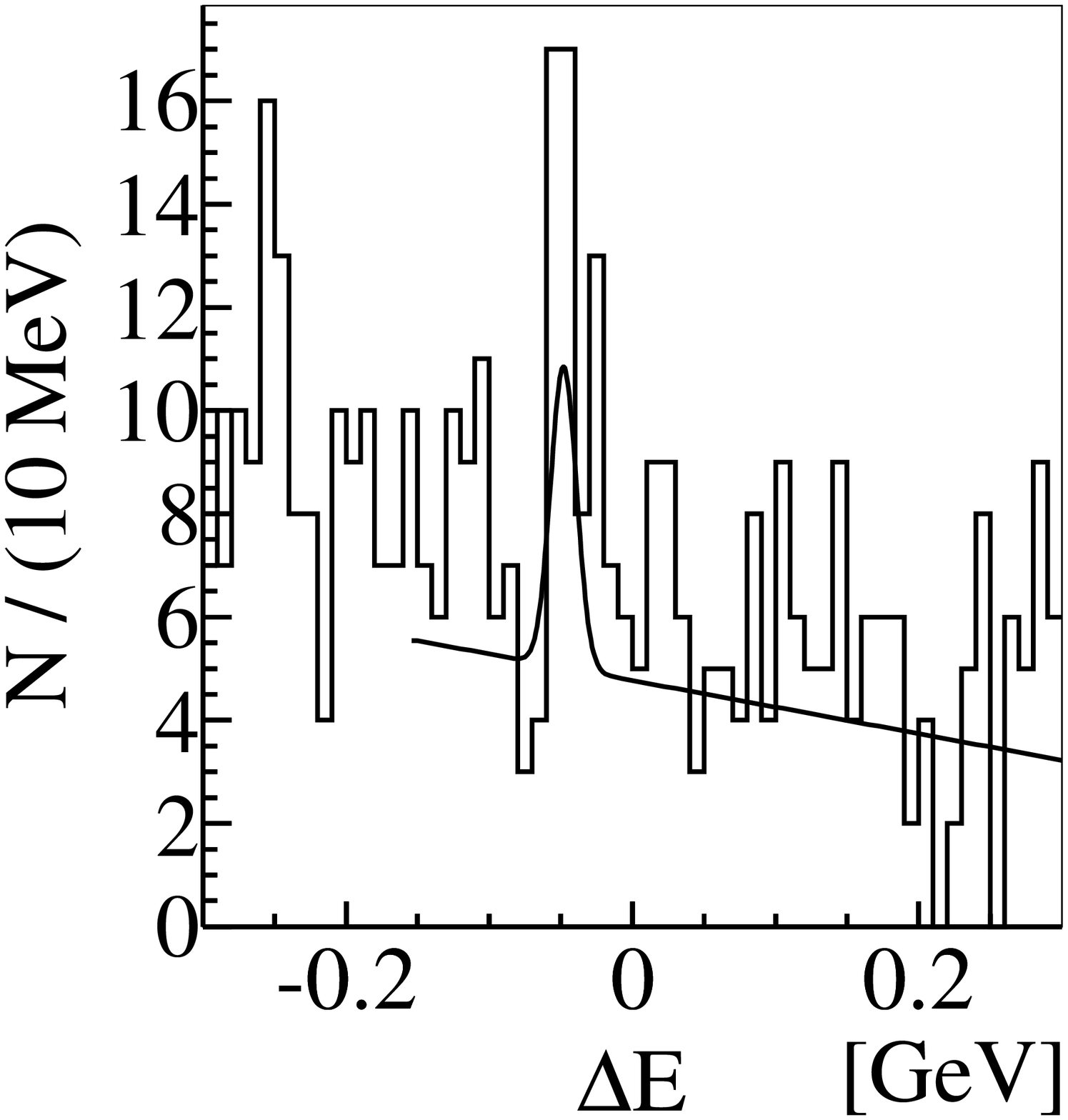}
   \includegraphics[width = 0.33\textwidth]{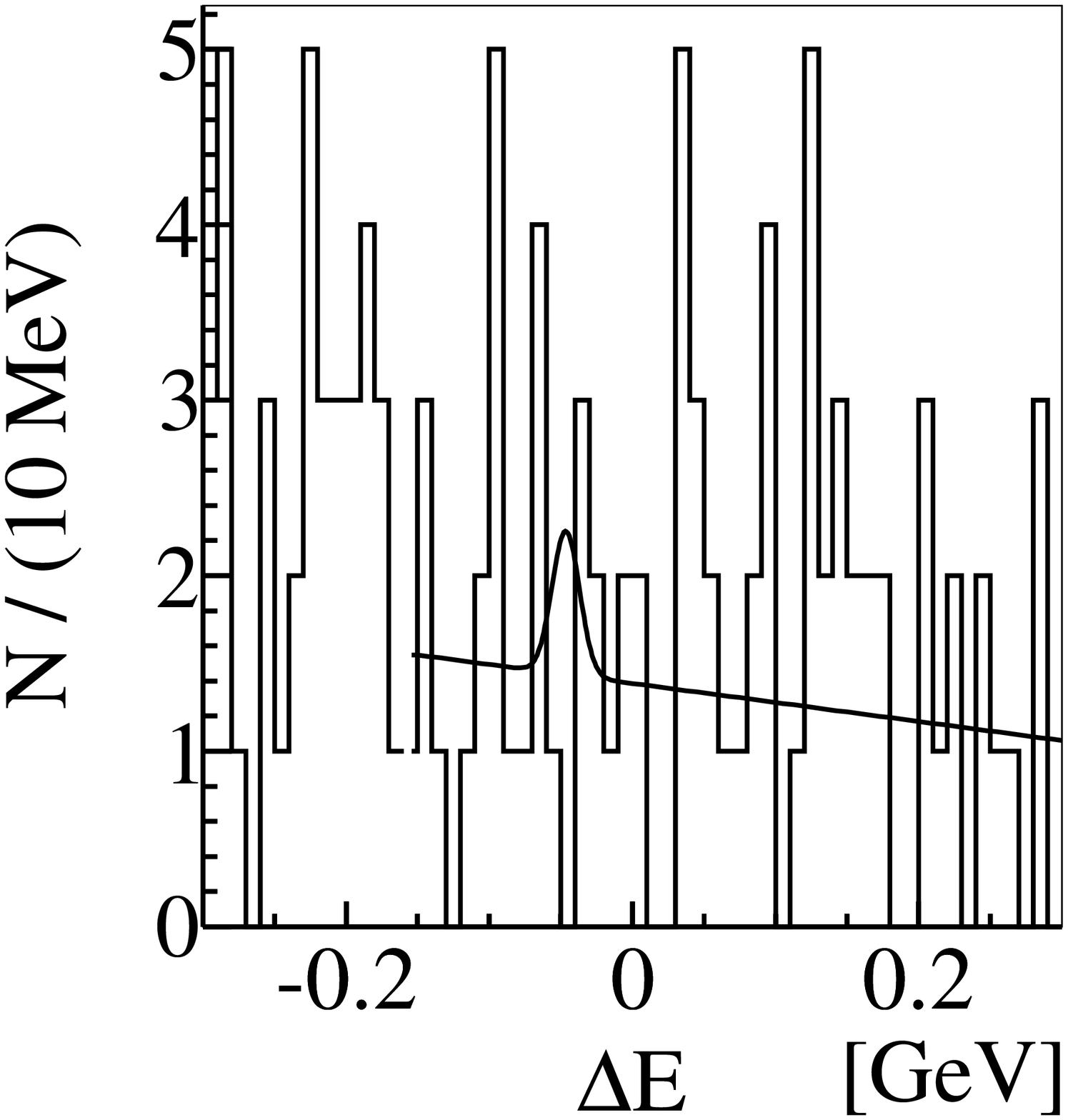}
   \put(-27,95){}
   \includegraphics[width = 0.33\textwidth]{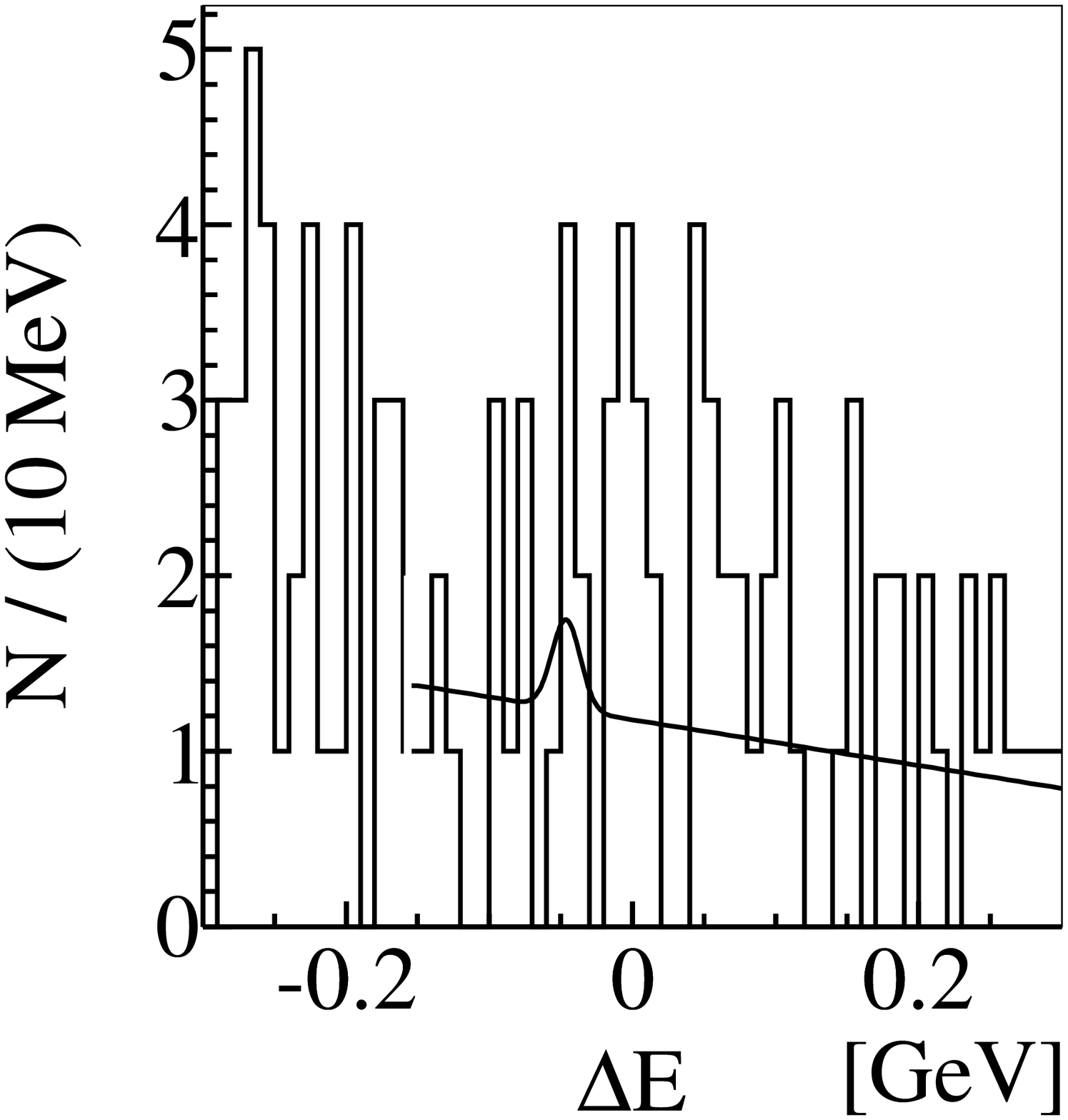}
   \caption{$\Mbc$ and $\DE$ projections for $\1$ followed by $\Lambda_c^+ \to p K^- \pi^+$ (left column), $\Lambda_c^+ \to p K_S^0$ (center column) and $\Lambda_c^+ \to \Lambda \pi^+$ (right column). $\Mbc$ spectra (top row) are for events in the $\Bs{*0} \Bsbar{*0}$ signal region ($-71 \mev < \DE< -23 \mev$) and $\DE$ spectra (bottom row) of the $\Lambda_c^+ \overline{\Lambda} \pi^-$ combinations are for events in the $\Bs{*0} \Bsbar{*0}$ signal region ($5.405 \gev/c^2 < \Mbc < 5.427 \gev/c^2$). The selection requirements and the fit are described in the text.}
   \label{Mbc_DE}
\end{figure}

The fit gives a $\1$ branching fraction of $(3.6 \pm 1.1) \times 10^{-4}$, which corresponds to a signal yield of $(20.3 \pm 6.1)$ events for the $\Lambda_c^+ \to p K^- \pi^+$ subchannel. In the $\Lambda_c^+ \to p K_S^0$ and $\Lambda_c^+ \to \Lambda \pi^+$ subchannels we expect a number of events compatible with zero and the fit results confirm these negligible yields within errors. The statistical significance of the observed signal is $4.4 \sigma$, which is calculated as $\sqrt{-2 \ln(L_0/L)}$, where $L_0$ and {\it L} are the likelihoods with the branching fixed at zero and at the best-fit value, respectively. This result provides the first evidence of a baryonic $\Bs{0}$ decay. Figure~\ref{Mbc_DE} shows the one-dimensional $\Mbc$ and $\DE$ projections for $\Bs{0}$ candidates from the $\Bs{*0} \Bsbar{*0}$ signal region ($5.405 \gev/c^2 < \Mbc < 5.427 \gev/c^2$, $-71 \mev < \DE< -23 \mev$).

For the systematic error calculation, we change the fixed signal parameters of the fit, reconstruction efficiencies, and fractions $f_C$ within their uncertainties, which gives a contribution of $+1.1 \atop -1.2$\%, and change the region excluded from the fit to avoid possible reflections, which results in a $+0.4 \atop -0.3$\% uncertainty. None of these fit variations lower the signal significance within the rounding accuracy. We also include a 0.35\% per track error to account for reconstruction uncertainties, a correlated systematic error of 2\% per \p{} and 1\% per \pion{} or \K{} to account for the particle identification efficiency, an uncertainty of $+0.0 \atop -6.8$\% due to the difference between data and Monte Carlo for tracks displaced from the IP and errors for all variables that enter into Eq.~(\ref{normalization}). Uncertainties from all sources are summarized in Table~\ref{syst} and are summed in quadrature.

Finally, we obtain the branching fraction:
$$
{\cal B}(\1) = \rm \left(3.6 \pm 1.1[stat.]~^{+0.3}_{-0.5}[syst.] \pm 0.9[\Lambda_c^+] \pm 0.7[N_{\Bsbar{0}}] \right) \times 10^{-4},
$$
where the uncertainties due to the $\Lambda_c^+$ absolute branching fractions~\cite{PDG} and total number of $\Bs{0}$-mesons are shown separately. The $B^- \to \Lambda_c^+ \overline{p} \pi^-$ mode, which represents a similar decay channel in the $B_u$-meson sector, has a branching fraction of $\left(2.8 \pm 0.8 \right) \times 10^{-4}$~\cite{PDG}.

\begin{table}[t]
   \renewcommand{\arraystretch}{1.2}
   \caption{Systematic uncertainties on ${\cal B}(\1)$}
   \centering
   \begin{tabular}{|l|c|}
      \hline
      \multicolumn{1}{|c|}{\multirow{2}{*}{Source}} & Relative \\
       & uncertainty,\% \\
      \hline
      \hline
      Fit parameters & $+1.1 \atop -1.2$ \\
      Cutoff & $+0.4 \atop -0.3$ \\
      Tracking efficiency & $\pm 2.1$ \\
      Particle identification & $\pm 8.0$ \\
      Displaced tracks & $+0.0 \atop -6.8$ \\
      $\Br{K_S^0 \to \pi^+ \pi^-}$ & $\pm 0.1$ \\
      $\Br{\Lambda \to p \pi^-}$ & $\pm 0.8$ \\
      \hline
      Total & $+8.4 \atop -12.8$ \\
      \hline
   \end{tabular}
   \label{syst}
\end{table}

To study the observed shapes of signal and background, we use only the $\Lambda_c^+ \to p K^- \pi^+$ subchannel as it contains the only portion of the signal. First, we examine distributions in the $\Lambda_c^+$ sidebands from $20 \mev/c^2 < |M \left(pK^- \pi^+ \right)-m_{\Lambda_c^+}| < 50 \mev/c^2$ and find no peaking structures in the signal area. Monte Carlo samples of known $\Upsilon(5S)$ decays that have six times the statistics of the dataset are analyzed using the same reconstruction procedure and requirements as described above. No hints for peaking structures in the signal $\Mbc$ and $\DE$ variables are seen. Finally, we check $\Upsilon(5S) \to B^{(*)} \overline{B}{}^{(*)} (\pi)$ processes~\cite{PDG} into which the $B^0$ decays to $\Lambda_c^+ \overline{\Lambda} \pi^-$. About 120000 signal events are generated and then analyzed using the same reconstruction procedure and requirements as described above. We find no peaking structures in the signal region, while a significant background contribution in the $\DE < -200 \mev$ region is seen. We conclude that the signal peak stems indeed from $\1$ decays.

\begin{figure}[t]
   \centering
   \includegraphics[width = 0.345\textwidth]{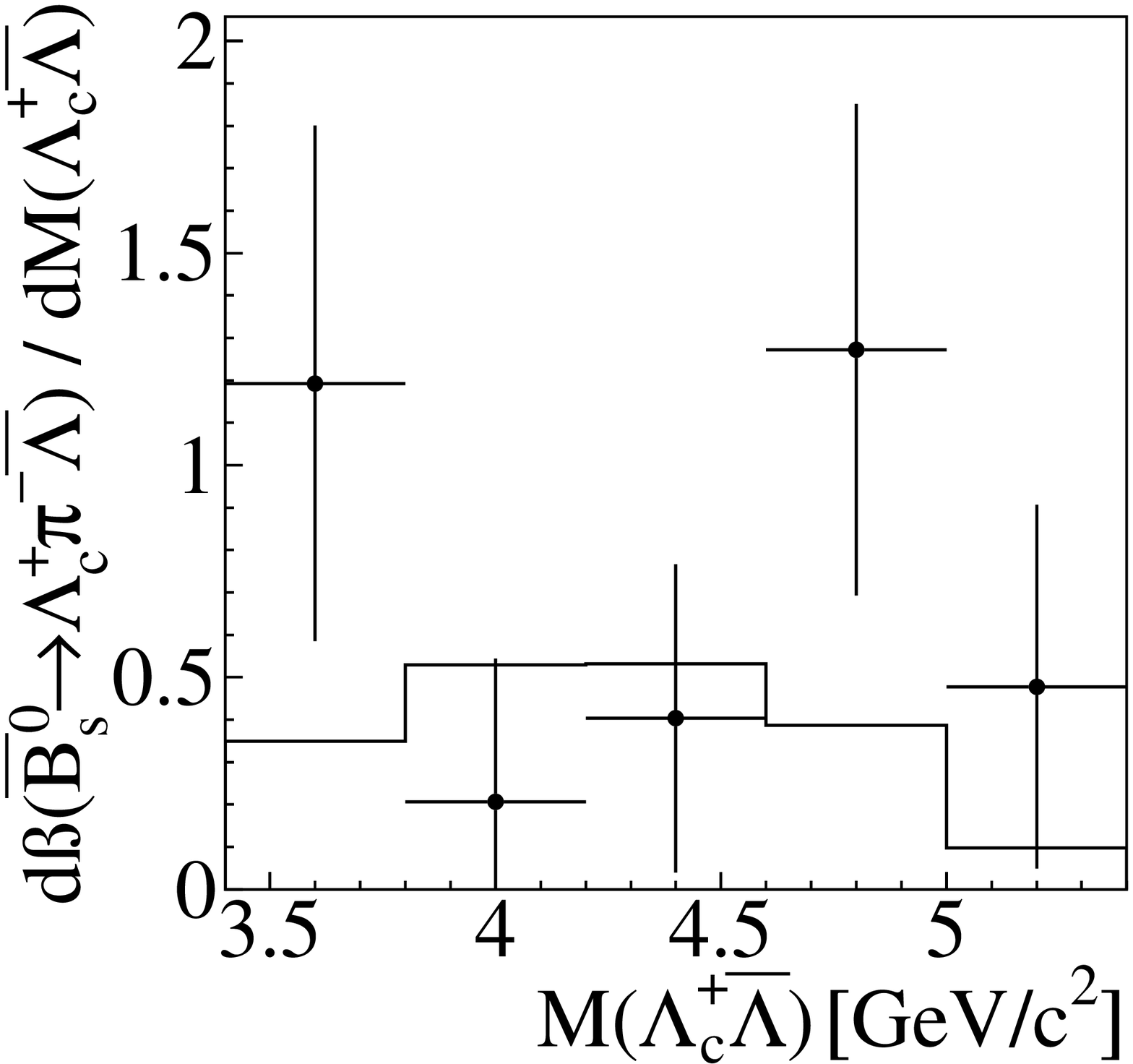}
   \put(-25,90){\Large(a)}
   \includegraphics[width = 0.345\textwidth]{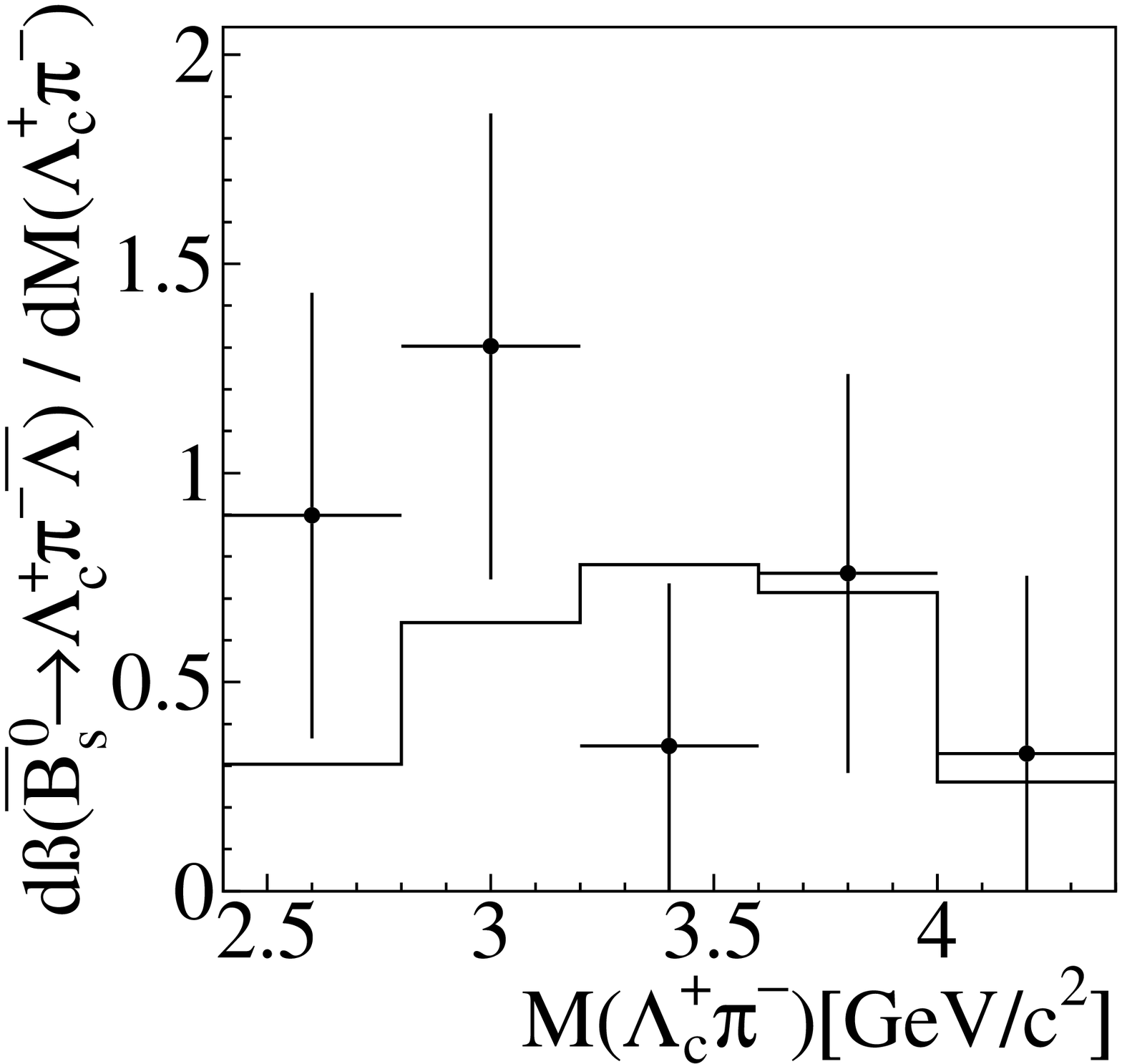}
   \put(-25,90){\Large(b)}
   \includegraphics[width = 0.345\textwidth]{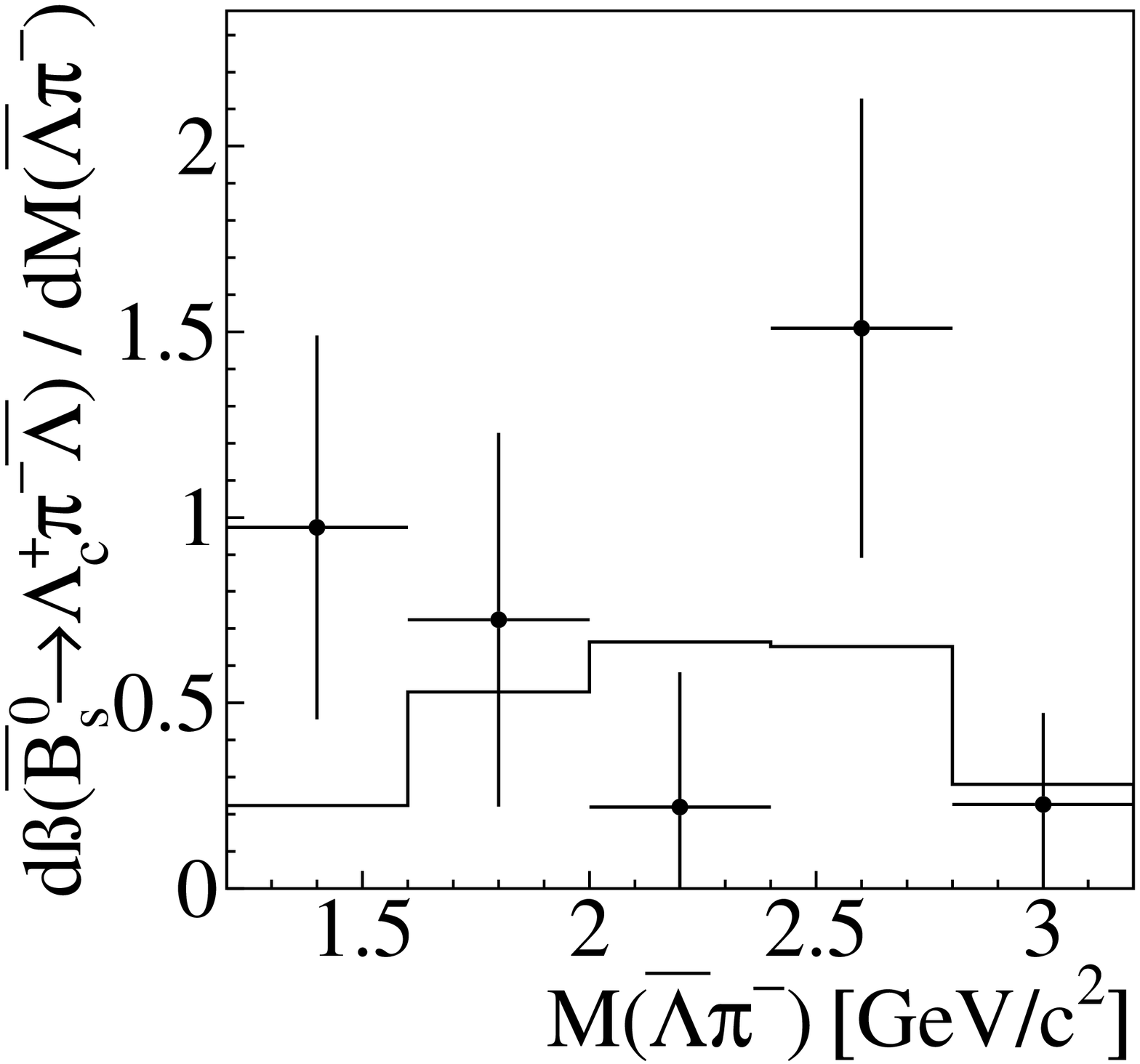}
   \put(-25,90){\Large(c)}
   \caption{Differential branching fraction as a function of (a) $M \left(\Lambda_c^+ \overline{\Lambda} \right)$, (b) $M \left(\Lambda_c^+ \pi^- \right)$, (c) $M \left(\overline{\Lambda} \pi^- \right)$. Points are the data; the solid histogram is the result of the phase-space fit. The vertical axis unit is $10^{-4} / (400 \mev/c^2)$.}
   \label{dibaryon}
\end{figure}

To investigate the possibility of a threshold enhancement, which is common in baryonic decays of $B_{u,d}$, including the related decay $B^- \to \Lambda_c^+ \overline{p} \pi^-$~\cite{Gabyshev}, we extract the signal yield in baryon-antibaryon mass bins and, after total reconstruction efficiency corrections, obtain differential branching fractions as a function of $M \left(\Lambda_c^+ \overline{\Lambda} \right)$; these are shown in Fig.~\ref{dibaryon}(a). A fit with a phase-space Monte Carlo simulated distribution with floating normalization gives a statistical compatibility of 19\%. We repeat the same procedure for other two-particle invariant masses: $\Lambda_c^+ \pi^-$ [see Fig.~\ref{dibaryon}(b)] and $\overline{\Lambda} \pi^-$ [see Fig.~\ref{dibaryon}(c)], finding 41\% and 22\% compatibility with the phase-space hypothesis. The present statistical precision is not sufficient to investigate the presence of a possible near-threshold effect.

\section{Summary}

In conclusion, we report the first evidence for the $\1$ decay and measure its branching fraction to be $\left(3.6 \pm 1.1[{\rm stat.}] {+0.3 \atop -0.5}[{\rm syst.}] \pm 0.9[\Lambda_c^+] \pm \right.$ $\left. 0.7[N_{\Bsbar{0}}] \right) \times 10^{-4}$ with a $4.4 \sigma$ significance, including systematics. The observed $\1$ process represents the first instance of a $\Bs{}$ baryonic decay.

We thank the KEKB group for the excellent operation of the accelerator, the KEK cryogenics group for the efficient operation of the solenoid, and the KEK computer group and the National Institute of Informatics for valuable computing and SINET4 network support.  We acknowledge support from the Ministry of Education, Culture, Sports, Science, and Technology (MEXT) of Japan, the Japan Society for the Promotion of Science (JSPS), and the Tau-Lepton Physics Research Center of Nagoya University; the Australian Research Council and the Australian Department of Industry, Innovation, Science and Research; the National Natural Science Foundation of China under contract Nos.~10575109, 10775142, 10875115 and 10825524; the Ministry of Education, Youth and Sports of the Czech Republic under contract Nos.~LA10033 and MSM0021620859; the Department of Science and Technology of India; the BK21 and WCU program of the Ministry Education Science and Technology, National Research Foundation of Korea, and NSDC of the Korea Institute of Science and Technology Information; the Polish Ministry of Science and Higher Education; the Ministry of Education and Science of the Russian Federation and the Russian Federal Agency for Atomic Energy; the Slovenian Research Agency;  the Swiss National Science Foundation; the National Science Council and the Ministry of Education of Taiwan; and the U.S. Department of Energy and the National Science Foundation. This work is supported by a Grant-in-Aid from MEXT for Science Research in a Priority Area (``New Development of Flavor Physics''), and from JSPS for Creative Scientific Research (``Evolution of Tau-lepton Physics'').


\begin{thebibliography}{99}
\bibitem{CC} Charge-conjugate modes are implicitly included throughout this Letter.
\end{thebibliography}
\end{document}